\documentclass[12pt]{article}

\usepackage{amsmath}
\usepackage{graphicx,psfrag,epsf}
\usepackage{enumerate}
\usepackage{graphicx}
\usepackage{amsmath, amsfonts, amssymb, mathrsfs, rotating, setspace, amsthm}
\usepackage{textcomp}
\usepackage{enumitem}
\usepackage[utf8]{inputenc}
\usepackage[margin=1in]{geometry}
\usepackage{url} 
\usepackage{algorithm,algorithmic}
\usepackage{natbib}
\usepackage{xcolor}
\usepackage{bbm}
\usepackage{rotating}
\usepackage{booktabs}
\usepackage{caption}
\usepackage{subcaption}
\usepackage{hyperref}
\usepackage{url}
\usepackage{natbib}
\usepackage{breakurl}
\usepackage{tablefootnote}
\usepackage{booktabs,caption}
\usepackage{tikz}
\usepackage{soul}
\usepackage{xcolor}

\theoremstyle{definition}
\newtheorem*{definition}{Definition}

\usetikzlibrary{positioning}
\usetikzlibrary{arrows}

\usepackage[flushleft]{threeparttable}

\newcommand{\blind}{1}

\begin{document}

\def\spacingset#1{\renewcommand{\baselinestretch}%
{#1}\small\normalsize} \spacingset{1}


\if1\blind
{
  \title{\bf The impact of coarsening an exposure on partial identifiability in instrumental variable settings}
  \author{Erin E. Gabriel$^1$\thanks{EEG is partially supported Novo Nordisk foundation NNF22OC0076595} \hspace{.2cm}\\
   Michael C. Sachs$^1$\\
   Arvid Sj\"olander$^2$\\
\footnotesize 1: Section of Biostatistics, Department of Public Health, University of Copenhagen, Denmark\\
\footnotesize 2: Department of Medical Epidemiology and Biostatistics, Karolinska Institutet, Stockholm, Sweden \\
corresponding author: erin.gabriel@sund.ku.dk}
  \maketitle
} \fi 

\if0\blind
{
  \bigskip
  \bigskip
  \bigskip
  \begin{center}
    {\LARGE\bf The impact of coarsening an exposure on partial identifiability in instrumental variable settings}
\end{center}
  \medskip
} \fi

\bigskip

\begin{abstract}
In instrumental variable (IV) settings, such as in imperfect randomized trials and observational studies with Mendelian randomization, one may encounter a continuous exposure, the causal effect of which is not of true interest.  Instead, scientific interest may lie in a coarsened version of this exposure. Although there is a lengthy literature on the impact of coarsening of an exposure with several works focusing specifically on IV settings, all methods proposed in this literature require parametric assumptions. Instead, just as in the standard IV setting, one can consider partial identification via bounds making no parametric assumptions. This was first pointed out in Alexander Balke's PhD dissertation. We extend and clarify his work and derive novel bounds in several settings, including for a three-level IV, which will most likely be the case in Mendelian randomization. We demonstrate our findings in two real data examples, a randomized trial for peanut allergy in infants and a Mendelian randomization setting investigating the effect of homocysteine on cardiovascular disease.


\end{abstract}

\newpage
\spacingset{2}
\vspace{-1cm}

\section{Introduction}  
We are interested in settings with an instrumental variable (IV), where the exposure, potentially unintentionally, is truly continuous but only the causal effect of a particular, potentially coarsened version, of the exposure is of interest. For example, in a randomized trial one might be randomized to take 14 doses of an experimental treatment versus placebo, and this may result in a realized exposure of 15 levels, anything between taking no doses and taking all 14 doses. When the intervention is, for example, a liquid or powder that needs to be measured out instead of a pill, this can lead to a nearly continuous realized exposure that is only bounded if the intervention is experimental and, therefore, is unavailable beyond the fixed quantity provided in the study. Often, this level of detail in the realized exposure is not even measured. Rather, only coarsened versions of the exposure are measured or of interest. There is a large literature on the topic of coarsening exposures in causal inference. Notably, \citet{stitelman2010impact} clearly outlines the population level assumptions required for the average treatment effect (ATE) based on a dichotomized exposure to have a meaningful causal interpretation. 

A number of papers have considered IV analysis under coarsened exposure. For example, \citet{tudball2021mendelian}, and \citet{howe2022interpreting} discuss coarsening in the Mendelian randomization setting, under linear structural equation models, allowing for both an improper IV and an ill-defining exposure. \citet{marshall2016coarsening} and \citet{angrist1995two}, Section 3.1., consider the bias and inconsistency imposed by coarsening under the monotonicity assumption that is typically made when using two stage least squares estimation. However, for all methods proposed in these papers, in order to identify and estimate the population causal effect, additional parametric assumptions beyond the standard IV assumptions are needed. 

Instead of making such assumptions, or in an additional sensitivity analysis, one can use nonparametric bounds to give a range for the possible causal effect. Chapter 6 of \citet{balke1995thesis} considers the type of noncompliance that may result in a realized continuous exposure in a randomized trial, and bounds under two different settings are provided. In the first setting, there are two levels of interest in the realized continuous exposure, the originally intended levels under perfect compliance and a third level that includes all other realized levels combined. Balke leaves the relationship between the third combined level and the outcome undefined and does not assume that the treatment assignment is a proper IV for that combined level. He also considers the setting where there are three levels, all of which are potentially of interest, and the treatment assignment is a proper IV for each of these three levels. We believe that these results from Balke's thesis were never published outside his dissertation, although they were mentioned in \citet{Balke97}, and so we reproduce them in this work. As noted in \citet{Balke97}, when gathering all but two levels of the exposure into a third level that is not of interest, and for which the treatment assignment may not be a proper IV, one arrives remarkably back at the classic IV bounds with a two-level well-defining exposure. Although implied, \citet{balke1995thesis} does not give a clear definition of the levels of interest that allow for these bounds to be applied, i.e., two ``well-defining'' levels and a third potentially ``ill-defining'' level. Additionally, \citet{balke1995thesis} implies that an ill-defining level always results in the lack of conditional independence between the IV and the outcome for that level. Finally, although three-level IVs are often used in Mendelian randomization settings, to our knowledge, bounds have never been derived in these settings allowing for a realized continuous exposure that is not of true interest. 

In this paper, we clearly define what is required for a potentially coarsened exposure level to lead to a ``well-defined" outcome, with anything not meeting these requirements leading to something ``ill-defined''.  We prove that an IV is not necessarily improper for a coarsened exposure, even if there are ill-defining levels. However, we also find that in settings where the coarsened third level of the exposure leads to a well-defined outcome, but is not of interest, and the IV has a direct effect on the outcome for that level, the tight and valid bounds are the same as those derived in \citet{balke1995thesis}. We also derive novel bounds in several new settings, including settings with only one well-defined outcome level and a two-level IV, and various coarsened exposures within a three-level IV setting (Mendelian randomization). We compare the novel bounds to previously published bounds and each other and find that the pattern found in \citet{balke1995thesis} and extended here persists in the three-level IV setting. Finally, we illustrate our findings in two real data settings: the \citet{du2015randomized} trial in peanut allergies and an observational study with Mendelian randomization to investigate the effect of homocysteine on cardiovascular disease \citep{meleady_thermolabile_2003}.

\section{Notation and preliminaries}
\begin{figure}[ht]
\centering
\begin{tikzpicture}
\node [draw, dashed, circle] (uh) at (3,1) {$\boldsymbol{U}$};
\node (z) at (0,0) {$Z$};
\node(x) at (2,0) {$X^*$};
\node (y) at (4,0) {$Y$};
\draw[-latex] (z) -- (x);
\draw[-latex] (x) -- (y);
\draw[-latex] (uh) -- (x);
\draw[-latex] (uh) -- (y);
\end{tikzpicture}
\caption{Causal diagrams in the IV setting \label{IVdag1}}
\end{figure}
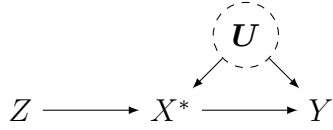

Let $Z$ be a two- or three-level IV, e.g., a randomized treatment assignment or a genetic polymorphism in a Mendelian randomization study. Let $X^*$ be the realized exposure under possible noncompliance, which may be continuous, let $Y$ be the binary outcome, and let $\boldsymbol{U}$ be the set of unmeasured confounders of $X^*$ and $Y$ with arbitrary distribution and dimension. The classic IV setting is illustrated by the directed acyclic graph (DAG) in Figure \ref{IVdag1}.

The DAG translates to the following structural equation model:
\begin{eqnarray}
z &=& g_{Z}(\epsilon_{z}) \\ \nonumber
x^* &=& g_{X^*}(\boldsymbol{u}, z, \epsilon_{x^*}) \\ \label{seqs0}
y &=& g_{Y}(\boldsymbol{u}, x^*, \epsilon_y^*), \nonumber
\end{eqnarray}
where the error terms $(\epsilon_{z},\epsilon_{x^*},\epsilon_y)$ are independent of each other, and of $\boldsymbol{U}$. This is not an assumption, but can be made to hold by absorbing all common causes of $(Z,X^*,Y)$ into $\boldsymbol{U}$. Let $X$ be the coarsened exposure, for which the sample space is a partition of the sample space of $X^*$ as defined by \citet{heitjan1991ignorability}. Since $X$ is by definition a deterministic function of $X^*$, we may represent $X$ by an additional equation $x=g_X(x^*).$

The estimands we consider are counterfactual probabilities $\psi_{x'}=p\{Y(X=x')=1\}$ and differences thereof of the form $\theta_{x'x} =p\{Y(X=x')=1\}-p\{Y(X=x)=1\}$, comparing two different levels for a potentially multi-level coarsened exposure.

\subsection{Well-defined potential outcome under coarsened exposure}
Consider now the setting where we know that the true realized exposure is not binary, but we are interested in a binary exposure. For example, $X=1$ if $X^* \in \{1,2\}$ and $X=0$ if $X^*=0$. Note that the counterfactual $Y(X=1)$ may be ill-defined in the sense that the intervention to set $X=1$ for everyone does not specify whether to set $X^*$ to 1 or 2, and the effect may differ across these two exposure levels. 

\begin{definition}
We say that the counterfactual $Y(X=x)$, and thus also the probability under the coarsened exposure, $p\{Y(X=x) = y\}$, is \emph{well-defined} for a particular level of the coarsened exposure $x$ if 
\begin{eqnarray*}
f_Y(x, u,\epsilon_Y) =g_Y(x^*,u,\epsilon_Y^*)=g_Y(x^{*'},u,\epsilon_Y^*)\textrm{ for all }(x^*,x^{*'})\textrm{ such that }g_X(x^*)=g_X(x^{*'})=x.
\end{eqnarray*}
\end{definition}
Essentially, this definition removes any ambiguity in the counterfactual expressions $Y(X=x)$ and  $p\{Y(X=x)=y\}$ by demanding that different interventions on $X^*$ give identical effects on $Y$ if they all result in $X$ having the value $x$, i.e. $Y(x)=Y(x^*)$.  In the example above $Y(X=0)$, is clearly well-defined as $X=0$ if and only if $X^*=0$, while, simultaneously, $Y(X=1)$ may be ill-defined. As the shorthand for the outcome, $Y$ being well-defined or ill-defined under the coarsening of $X^*$ to $X$ for a given level $x$, we will refer to the levels of the coarsened exposure as well-defining or ill-defining.

Our well-defined definition can be seen as the individual level counterpart of the population level assumption 2 of \citet{stitelman2010impact}. However, if, as in the above example, $X=x$ if and only if $X^*=x^*$, which can be seen as the individual level version of  Assumption 1 of \citet{stitelman2010impact}, then $Y(x)=Y(x^*)$ also holds. Thus, $Y(x)=Y(x^*)$ can be seen as the individual-level version of both population assumptions in \citet{stitelman2010impact}. It is of note that $X=x$ if $X^*=x^*$ would also induce $Y(x)=Y(x^*)$, but this would imply refinement rather than a coarsening of the exposure. 

In order to make progress for partial identification, we need to assume that there are at least some levels of $Y(X=x)$ for the coarsened exposure $X$ that are well-defined.  We need not assume that all levels of $X$ are well-defining, nor that the same NPSEM as given in the classic IV setting hold for the potentially ill-defining levels of $X$. When all coarsened levels of the exposure are well-defining, and we are in the setting of Figure \ref{IVdag1}, we will consider the NPSEM
\begin{eqnarray}
z &=& g_{Z}(\epsilon_{z}) \\ \nonumber
x^* &=& g_{X^*}(\boldsymbol{u}, z, \epsilon_{x^*}) \\ \nonumber
y &=& f_{Y}(\boldsymbol{u}, g_X(x^*), \epsilon_y)=f_{Y}(\boldsymbol{u}, x, \epsilon_y)\\ \label{seqs}
x &=& g_X(x^*)  \nonumber
\end{eqnarray}

\subsection{Validity of an IV}
 There are several characterizations of a `valid' IV in the literature. Often, an IV $Z$ is said to be valid \citep{didelez2007mendelian} with respect to an exposure $X^*$ and an outcome $Y$ if it satisfies the following conditions: 
\begin{enumerate}
\item $Z$ is independent of $U$.
\item $Z$ is associated with $X^*$.
\item $Z$ and $Y$ are conditionally independent given $(X^*,U)$.
\end{enumerate}
An IV that violates conditions 1 or 3 is often referred to as a contaminated IV, while the violation of condition 2 is often called a useless IV. 

\subsection{Previous bounds}
In settings with three well-defining coarsened levels of exposure $X \in\{x,x',x''\}$, all of which are possibly of interest and based on all of which $Z$ remains a proper IV, but making no additional assumptions beyond the NPSE models encoded by the DAG in \ref{IVdag1}, the bounds for $ \theta_{x'x}$ given in terms of the value $x$, $x'$ and $x''$ such that $x'' \notin \{x,x'\}$ are as follows, where $p_{xy\cdot z} = p(X=x, Y=y|Z=z)$.

\begin{eqnarray}
 \theta_{x'x} &\geq& \mbox{max} \left. \begin{cases}   
   -1 + p_{x0\cdot 1} + p_{x'1\cdot 1},\\ 
    -1 + p_{x0\cdot 1} + p_{x'1\cdot 0} \\
   -1 + p_{x0\cdot 0} + p_{x'1\cdot 0},\\ 
 -1 + p_{x0\cdot 0} + p_{x'1\cdot 1},\\ 
   -2 + 2p_{x0\cdot 0} + p_{x1\cdot 1} + p_{x'1\cdot 0} + p_{x'1\cdot 1},\\ 
   -2 + p_{x0\cdot 0} + p_{x0\cdot 1} + p_{x'0\cdot 0} + 2p_{x'1\cdot 1},\\ 
     -2 + 2p_{x0\cdot 1} + p_{x1\cdot 0} + p_{x'1\cdot 0} + p_{x'1\cdot 1},\\ 
   -2 + p_{x0\cdot 0} + p_{x0\cdot 1} + p_{x'0\cdot 1} + 2p_{x'1\cdot 0},\\ 
     -2 + p_{x0\cdot 1} + p_{x'1\cdot 0} + p_{x''1\cdot 0} + p_{x''0\cdot 1} + p_{x0\cdot 0} + p_{x'1\cdot 1},\\ 
   -2 + p_{x0\cdot 0} + p_{x'1\cdot 1} + p_{x''0\cdot 0} + p_{x''1\cdot 1} + p_{x0\cdot 1} + p_{x'1\cdot 0}
   \end{cases}  \label{lower1} \right\} 
\end{eqnarray}

\begin{eqnarray}
 \theta_{x'x} &\leq& \mbox{min} \left. \begin{cases}   1 - p_{x'0\cdot 1} - p_{x1\cdot 0},\\ 
    1 - p_{x'0\cdot 1} - p_{x1\cdot 1},\\ 
   1 - p_{x'0\cdot 0} - p_{x1\cdot 0},\\ 
      1 - p_{x'0\cdot 0} - p_{x1\cdot 1},\\ 
   2 - 2p_{x'0\cdot 1} - p_{x1\cdot 0} - p_{x1\cdot 1} - p_{x'1\cdot 0},\\ 
   2 - p_{x0\cdot 1} - p_{x'0\cdot 0} - p_{x'0\cdot 1} - 2p_{x1\cdot 0},\\ 
 2 - 2p_{x'0\cdot 0} - p_{x1\cdot 0} - p_{x1\cdot 1} - p_{x'1\cdot 1},\\ 
   2 - p_{x0\cdot 0} - p_{x'0\cdot 0} - p_{x'0\cdot 1} - 2p_{x1\cdot 1},\\
   2 - p_{x1\cdot 0} - p_{x'0\cdot 1} - p_{21\cdot 0} - p_{20\cdot 1} - p_{x1\cdot 1} - p_{x'0\cdot 0},\\ 
   2 - p_{x1\cdot 1} - p_{x'0\cdot 0} - p_{20\cdot 0} - p_{21\cdot 1} - p_{x1\cdot 0} - p_{x'0\cdot 1}  
\end{cases}   \label{upper1} \right\} 
\end{eqnarray}

 Although they are displayed differently, these are equivalent to the bounds given in \citet{balke1995thesis} pages 128-129. Here we are assuming that the realized exposure $X^*$ can be coarsened into exactly three well-defining levels. As this is unlikely to be the case when the exposure is realized, via noncompliance, to be continuous and then coarsened, \citet{balke1995thesis} point out that we need not make this assumption. 
 
In this setting, one can collect all uninteresting levels of $X^*$ into one potentially ill-defining level $x^{m}$, and two well-defining levels of interest $X \in\{x,x'\}$. Although this is difficult to depict in a DAG, the resulting NPSEM are as follows: 
\begin{eqnarray} \label{npsemill}
z &=& g_{Z}(\epsilon_{z}) \\ \nonumber
x^* &=& g_{X}(\boldsymbol{u}, z, \epsilon_{x^*}) \\ \nonumber
x &=& g_X(x^*) \\ \nonumber
y &=& \begin{cases} 
f_{Y}(\boldsymbol{u}, x, \epsilon_y) \mbox{ if } x \in \{x, x'\} \\ 
\mbox{undefined if } x = x^m 
\end{cases}.
\end{eqnarray}
As pointed out in \citet{Balke97}, the resulting bounds for the $\theta_{x'x}$ are `remarkably' the same IV bounds for a two-level well-defining exposure presented in that same work; note the bounds in \citet{Balke97} are given in the notation is $p_{yx\cdot z} = p(Y=y,X=x|Z=z)$. It is of note that these are the first eight terms of each set of the above bounds, \eqref{lower1} and \eqref{upper1}. 

In deriving the bounds for $\theta_{x'x}$ allowing for a possibly ill-defining third level $x^{m}$, \cite{balke1995thesis} seems to imply that coarsening an exposure and allowing for a potentially ill-defining third level also always allows for a contaminated IV for that level, i.e. an ill-defining level implies a contaminated IV. We show this is not inherently true below.

\section{Validity of an IV under a coarsened exposure}
Given a coarsening $X$, it may be perceived as natural to say that $Z$ remains a valid IV with respect to $X$ and $Y$ if it satisfies conditions 1-3 above, when $X^*$ is replaced by $X$. As condition 1 does not relate to $X^*$ (or $X$), we only consider 2 and 3 i.e.,\\
\\
2$^{\star}$. $Z$ is associated with $X^*$.\\
3$^{\star}$. $Z$ and $Y$ are conditionally independent, given $(X^*,U)$.\\
\\
These conditions may be satisfied even when $p\{Y(X=x)=y\}$ is ill-defined.\\ \noindent
Proof (by example): Suppose that $X\in\{0,1,2\}$ and define the coarsening $X=I(X^*>0)$. Combining the definition of $X$ with the assumption that condition 3 is satisfied gives that condition 3$^{\star}$ holds for $X=0$. To construct a scenario where it holds for $X=1$, note that 
\begin{eqnarray*}
p(Y|Z,U,X=1)&=&\sum_{x^*\in\{1,2\}}p(Y|Z,U,X^*=x^*)p(X^*=x^*|Z,U,X=1)\\
&=&\sum_{x^*\in\{1,2\}}p(Y|U,X^*=x^*)\frac{p(X^*=x^*|Z,U)}{\sum_{x^*\in\{1,2\}}p(X^*=x^*|Z,U)},
\end{eqnarray*}
where the second equality follows from the assumption that condition 3 is satisfied and the definition of $X$. The right hand side does not depend on $Z$, so that condition 3$^{\star}$ is satisfied, if $
\frac{p(X^*=1|Z,U)}{p(X^*=2|Z,U)}=g(U)$
for some function $g(U)$ that does not depend on $Z$. To additionally make condition $2^{\star}$ hold, define $
\frac{p(X^*=0|Z,U)}{p(X^*=2|Z,U)}=h(U,Z)$
for some function $h(U,Z)$ that depends on both $U$ and $Z$. It follows that 
$$
p(X^*=0|Z,U)=\frac{h(U,Z)}{1+g(U)+h(Z,U)} \mbox{, }
p(X^*=1|Z,U)=\frac{g(U)}{1+g(U)+h(Z,U)}
$$
so that 
\begin{eqnarray*}
p(X=0|Z=z)&=&E\left\{\frac{h(U,z)}{1+g(U)+h(z,U)}\Big | Z=z\right\}\\
&=&E\left\{\frac{h(U,z)}{1+g(U)+h(z,U)}\right\}\\
p(X=1|Z=z)&=&E\left\{\frac{g(U)}{1+g(U)+h(z,U)}\Big | Z=z\right\}\\
&=&E\left\{\frac{g(U)}{1+g(U)+h(z,U)}\right\},
\end{eqnarray*}
where the second set of equalities follow from the assumption that condition 1 is satisfied. The right hand side of these can be made to depend on $z$, irrespective of $p(U)$, since $h(U,Z)$ is arbitrary. Thus, condition $2^{\star}$ can be made to hold. 

Note, we did not assume any form of $g_Y(X^*,U,\epsilon_Y)$. Thus, without violating the above proof we can specify $g_Y(X^*,U,\epsilon_Y)$ so that $p\{Y(X^*=1)=y\}\neq p\{Y(X^*=2)=y\}$. This makes $p\{Y(X=1)=y\}$ ill-defined, since the intervention $X=1$ may correspond to either $X^*=1$ or $X^*=2$, which give different counterfactual outcomes, but $Z$ is still a valid IV with regard to $X$. Thus, an ill-defining level of the exposure does not imply an improper IV given that outcome, nor does an improper IV imply an ill-defining exposure.  

As is evident from the NPSEM in \eqref{npsemill}, given the undefined NPSE for $Y$ when $X=x^{m}$, we cannot enforce that this function is not dependent on $Z$, i.e. enforce condition 3$^{\star}$. However, given a well-defining third level of $X$, allowing for an improper IV is possible, and thus, it is of interest to determine if the bounds for $\theta_{x'x}$ allowing for three well-defining levels, but a contaminated IV for one of those levels differ from those allowing for a third ill-defining level. 

\section{Novel bounds and results}
Assuming there are three well-defining levels of the exposure, but for  one level, which we will continue to call $x^{m}$, there is a direct effect of the IV $Z$ on the outcome $Y$, we derive tight symbolic bounds for $\theta_{x'x}$. Thus, assuming that the contrast of interest is between the other two levels. Although this is difficult to depict in a DAG, the resulting NPSEM is as follows:
\begin{eqnarray}\label{seqsvio3}
z &=& g_{Z}(\epsilon_{z}) \\ \nonumber
x^* &=& g_{X}(\boldsymbol{u}, z, \epsilon_{x^*}) \\\nonumber
x &=& g_X(x^*) \\ \nonumber
y &=& \begin{cases} 
f_{Y}(\boldsymbol{u}, x, \epsilon_y) \mbox{ if } x \in \{x, x'\} \\  
f_{Y}(\boldsymbol{u}, x, z, \epsilon_y) \mbox{ if } x = x^m
\end{cases}.
\end{eqnarray}
These NPSEM violate condition 3$^{\star}$. 

\noindent \textbf{Result 1}: For a two-level IV and two well-defining levels of the exposure $x$ and $x'$ and one potentially ill-defining level $x^{m}$, NPSEM \eqref{npsemill} or three well-defining levels $x$, $x'$ and $x^{m}$, for only one of which, $x^{m}$, there is a direct effect of the IV $Z$ on $Y$, NPSEM \eqref{seqsvio3} the resulting tight and valid bounds for $\theta_{x'x}$, which is not a contrast in $x^{m}$, are symbolically identical to the well known two-level IV bounds given in \citet{Balke97} (the first eight terms respectively in \eqref{lower1} and \eqref{upper1}). 

Proof of this result follows from the linearity of the problem when the contrast of interest does not involve the level with the contaminated IV $x^{m}$, in addition to the results of \citet{balke1995thesis}. When the contrast of interest does involve $x^{m}$, i.e., $\theta_{x^{m}x}$, the resulting target is no longer linear in the counterfactual response function probabilities, and thus the linear programming method of \citet{sachs2022general} cannot be used. Further discussion is provided in the supplementary materials in addition to the proof of Result 1. 

Remarkably, the resulting bounds do not involve $x^{m}$, even though it is a well-defining level. This result implies that a well-defining third level under which the IV has a direct effect on the outcome provides no additional information for narrowing the range of the bounds. Of course, contamination is only one way that an IV can be improper, and violations of 2$^{\star}$ would also be of interest. Unfortunately, the system of NPSEMs where $Z$ is d-separated from only one level of $X$ is not linear in the constraints, regardless if the contrast involves $x^{m}$, and thus, at least using the method of \citet{sachs2022general}, tight symbolic bounds cannot be calculated. 

It is possible that the coarsening of an exposure can result in not having enough well-defining levels for a contrast to be well-defined. Under the assumption that there is only one well-defining level of the coarsened exposure $X$, under which there is still a proper IV, we now consider bounds for $p\{Y(X=x)=1\}$. Here $x$ is the only assumed well-defining level of $X$ in the setting of the DAG in Figure \ref{IVdag1}, while all other levels of $X^*$ have been gathered into a possibly ill-defining level $x^{m}$. Here, we cannot give bounds for a contrast between levels of $X$, so instead, we bound the counterfactual probability $\psi_{x}=p\{Y(X=x)=1\}$. We also consider the setting where $x^{m}$ is well-defining, but for this level of exposure, the IV has a direct effect on the outcome. 

\noindent \textbf{Result 2}: For a two-level IV and only one well-defining level of $X$, $x$ and one potentially ill-defining level $x^{m}$, as in the NPSEM \eqref{npsemill} or two well-defining levels for one of which, $x^{m}$, the IV has a direct effect on the outcome, NPSEM \eqref{seqsvio3} the tight and valid bounds for $\psi_{x}$ are as follows: 
\[ 
 \mbox{max} \left. \begin{cases}   p_{x1\cdot 0},\\
  p_{x1\cdot 1} \end{cases} \right\} 
\leq \psi_{x}
 \leq \mbox{min} \left. \begin{cases}   1 - p_{x0\cdot 0},\\ 
   1 - p_{x0\cdot 1} \end{cases} \right\}.
 \]
Unlike in Result 1, where the bounds are only symbolically identical, here, as there are two levels of exposure in both settings, these bounds are the same both symbolically and numerically.

\citet{balke1995thesis} consider a two-level IV, likely most common in randomized settings, while three-level IVs are much more likely in Mendelian randomization settings. We, therefore, consider bounds for $\theta_{x'x}$ in settings with a three-level IV for a coarsened exposure with either two- or three well-defining levels and two- or three well-defining levels and an additional potentially ill-defining level $x^{m}$. Again, we also consider the setting where the IV has a direct effect on the outcome for one of the levels that is not involved in the contrast. The resulting bounds are too long to provide symbolically in the main text; they are given in the supplementary materials Section S1 for the three well-defining exposure levels. R code to implement the bounds in a given data set is also available as a supplementary file. The bounds for a three-level IV and two well-defining levels of exposure have been previously published in \citet{palmer2011nonparametric, jonzon2022accessible}. However, the derivation of the bounds in the six cases yielded the following results. 

\noindent \textbf{Result 3}: For a three-level IV and two well-defining levels of the coarsened exposure $X$, under DAG \ref{IVdag1}, NPSEM \eqref{seqs}, the tight and valid bounds for $\theta_{x'x}$ are symbolically identical to the tight and valid bounds allowing for a third potentially ill-defining level of the coarsened exposure, $X$, $x^{m}$, NPSEM \eqref{npsemill}, or for a well-defining third level for which there is a direct effect of the IV on the outcome, NPSEM \eqref{seqsvio3}. 

\noindent \textbf{Result 4}: For a three-level IV and three well-defining levels of the coarsened exposure $X$, under DAG \ref{IVdag1}, NPSEMs \eqref{seqs}, the tight and valid bounds for $\theta_{x'x}$ are symbolically identical to the valid and tight bounds allowing for a fourth potentially ill-defining level of the coarsened exposure, $X$, $x^{m}$, NPSEMs \eqref{npsemill}, or for a well-defining fourth level for which there is a direct effect of the IV on the outcome, NPSEMs \eqref{seqsvio3}.

Given Results 3 and 4, we know that the three-level IV bounds display the same pattern as the two-level IV bounds, i.e. they are the same for a two- or three-level well-defining exposure with or without the addition of an exposure level that is potentially ill-defining, or an exposure level for which the IV has a direct effect on the outcome. An outline of the Proof of Results 3 and 4 is given in the supplementary materials Section S2, following for format of the proof of Result 1. 

\section{Comparison of bounds}
As the tight and valid bounds for $\theta_{x'x}$ comparing any two levels within a well-defining three-level exposure for settings with a two-level always proper IV contain all terms in classic two-level IV bounds, it immediately follows that the two-level exposure bounds are always valid but possibly not tight in the setting of a well-defining three-level exposure. This is clear because a maximum can never be made smaller by adding additional values, nor can a minimum be made larger. 

As the tight and valid bounds for $\theta_{x'x}$ comparing the two levels within a well-defining two-level exposure and a two-level always proper IV are the same as the valid bounds allowing for an ill-defining third level, the tight and valid bounds for a well-defining three-level exposure are possibly invalid in settings with an ill-defining third level. To demonstrate this, we simply need to find a place where one of the two additional terms is the maximum in \eqref{lower1} or the minimum in \eqref{upper1}, as the first eight terms of each are the classic IV bounds. This is easily observed when $p_{x''1\cdot 0} = p_{x''0\cdot 1}=1$, which sets the second to last term in \eqref{lower1} to 0, while all other terms are $<0$ and the second to last term in \eqref{upper1} to 0, while all other terms $>0$. Thus, identifying $\theta_{x'x}=0$. This may seem like an extreme case, however, it shows that it is possible to obtain invalid bounds when assuming three well-defining levels of a coarsened exposure, when only two well-defining levels exist within a setting. 

Given Result 1, we have also shown that if there are three well-defining levels of the coarsened exposure, but via the level not involved in the contrast $\theta_{x'x}$ the IV has a direct effect on the outcome, we again arrive at the classic IV bounds for two well-defining levels. Thus, just as above, if one assumes a well-defining three-level exposure given which there is a valid two-level IV the bounds for $\theta_{x'x}$ are invalid if the IV is invalid, for even one of the levels of the exposure, even if that level is not referenced in the contrast $\theta_{x'x}$. However, if one uses the bounds allowing the level with the improper IV to be ill-defining they will be valid and tight, even if the level is actually well-defining. Results 3 and 4, allow for the same logic for a three-level IV, and an either two-level or three-level exposure of interest. 

Comparing the bounds in Result 2 to the bounds for $\psi_x$ given in \citet{Balke97}, we see that our bounds are the first two terms in the upper and lower bounds, respectively, of \citet{Balke97}. Again this implies that the bounds only assuming one well-defining level of the coarsened exposure and allowing for all other levels of $X^*$ to be gathered into a potentially ill-defining level are valid, but possibly not tight, for $\psi_x$ when more levels of the exposure are assumed to be well-defining or when more levels of the exposure block the path from the IV to the outcome. 

\section{Estimation}
So far we have displayed and discussed the bounds in terms of the true observable probabilities. These probabilities of course need to be estimated in practice. The bounds can be estimated nonparametrically by taking the proportion within observable groups. Although this may be very inefficient given that very few measurements may be observed with the well-defining levels, any form of estimation that might increase the effective sample size would require additional parametric assumptions, and thus threaten the validity of the results. Beyond this, for inference we suggest the nonparametric bootstrap for the contrast estimands as was suggested and demonstrated in \citet{gabriel2021nonparametric}. For the single probability estimands, which are more likely to be of interest when they lie on the boundary of the parameter space, we suggest the use of $m$-out-of-$n$ bootstrap as suggested by \citet{andrews2000inconsistency} and demonstrated in \citet{gabriel2023sharp}. We demonstrate both forms of bootstrap in the real data examples. 

\section{Motivating Data Examples}
\begin{figure}[ht]
\centering
\includegraphics[scale = .9]{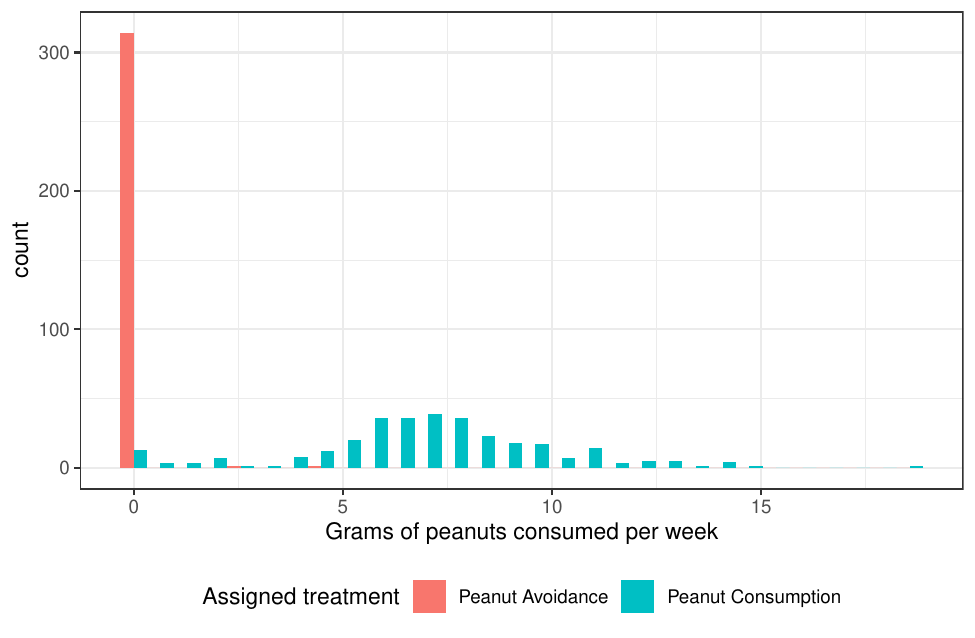}
\caption{Histogram showing the frequency of the number of grams of peanut products consumed per week, by treatment assigned. \label{peanutgramhist}}
\end{figure}

The results of the peanut allergy trial are presented in \citet{du2015randomized}. The publicly available trial data were downloaded from the Immune Tolerance Network TrialShare website on 2020-06-15 (\url{https://www.itntrialshare.org/}, study identifier: ITN032AD).
The trial had a two-armed randomization with one arm assigned to consume a certain amount of peanut powder per week and one arm assigned to avoid consuming peanut powder. However, the true realized exposure is likely continuous as the true amount of peanut powder ingested by young children is dependent not only on being given the powder, but also on how much of it is eaten rather than left on the plate, floor, or table in addition to many other elements that may change absorption once ingested. In this study, 640 participants between 4 months and 11 months of age were randomized to either consume peanuts or avoid peanuts until the age of 60 months. Those in the consumption arm were instructed to consume at least 6 grams per week. At the end of the study, the outcome allergy to peanuts was assessed using a oral food challenge. Compliance with the assigned intervention was assessed weekly by using a food frequency questionnaire, and by manual inspection of the infants' cribs for peanut crumbs in a randomly selected subset of participants. Thus the true treatment taken could be considered continuous as the observed number of grams of peanut products consumed during the study period. We used the adherence data to define the continuous variable grams of peanuts consumed per week, the distribution of which is shown by arm in histograms in Figure \ref{peanutgramhist}.

Based on this continuous version of the treatment taken, we consider a coarsened version of the exposure, a ternary treatment variable defined by categorizing the grams consumed per week into less than 0.2g, 0.2 to 6g, and 6g or more. Based on this coarsened treatment, we have the data shown in Table \ref{peanutgramtable}, which we will assume is sufficient to calculate bounds on the risk difference $p\{Y(X^* = <0.2\mbox{g}) = 1\} - p\{Y(X^* = 6\mbox{g or more}) = 1\}$. Interestingly, regardless of whether we assume that all three levels are well-defining, only the two levels are well-defining, or that the assignment may have a direct effect through the level 0.2 to 6g, we get the same numeric bounds: $-0.16$ to $0.16$ (with 95\% bootstrap confidence interval -0.20 to 0.21). 

Based on a coarsened binary treatment variable defined by categorizing grams consumed per week into less than 0.2g versus equal or greater than 0.2g, we calculate bounds on the risk of allergy in the ``no consumption" intervention $p\{Y(X < 0.2\mbox{g}) = 1\}$. Whether we assume that the equal or greater than 0.2g category is well-defining or not, or contaminated or not, we obtain the bounds on the risk of allergy of $0.15$ to $0.20$ with $m$ out of $n$ bootstrap confidence intervals $0.05$ to $0.29$. The $m$ in the $m$ out of $n$ bootstrap was selected according to the method proposed by \citet{bickel2008choice}.

\begin{table}[ht]
\caption{Counts and percents of coarsened treatment taken and outcomes, by assignment group, in the peanut allergy study \citep{du2015randomized} \label{peanutgramtable}}
\centering
\begin{tabular}{lrr|lrr}
\multicolumn{3}{r|}{Avoidance Arm ($n = 305$)} & \multicolumn{3}{c}{Consumption Arm ($n = 312$)} \\
Outcome & Tolerant & Allergic & & Tolerant & Allergic \\
 \hline
 Observed peanut exposure & & & & \\
 $<0.2$g & 255 (83.6\%) & 48 (15.7\%) & & 6 (2.0\%) & 6 (2.0\%) \\
 $0.2 - 6$g & 2 (0.7\%) & 0 (0.0\%) & & 84 (27.5\%) & 3 (1.0\%) \\
 $6$g or more & 0 (0.0\%) & 0 (0.0\%) & & 213 (69.8\%) & 0 (0.0\%) 
\end{tabular}
\end{table}

\citet{meleady_thermolabile_2003} report the results of an observational study designed to investigate the effect of homocysteine on cardiovascular disease using the 677CT polymorphism (rs1801133) in the Methylenetetrahydrofolate Reductase gene as an instrument. In this case the outcome is binary, the treatment has been made categorical in 6 levels, and the instrument is ternary (this polymorphism can take three possible genotype values). Summary data are available in Table 3 of \citet{meleady_thermolabile_2003}, and the relevant summary statistics have been reproduced in Table \ref{menddat}. 

\begin{table}[ht]
\caption{Data from the cardiovascular disease study reported in \citet{meleady_thermolabile_2003}. \label{menddat}}
\centering
\begin{tabular}{llll}
  \hline
  & \multicolumn{3}{c}{Cardiovascular disease absent} \\
Genotype & CC ($n = 665$) & CT ($n = 621$) & TT ($n = 172$)\\ 
  \hline
  Homocystine group ($\mu$mol/L) & & & \\
$<$9 & 164 (24.7\%) & 133 (21.4\%) & 16 (9.3\%) \\ 
  9–14.99 & 177 (26.6\%) & 164 (26.4\%) & 47 (27.3\%) \\ 
  15–20 & 11 (1.7\%) & 15 (2.4\%) & 9 (5.2\%) \\ 
  20–30 & 0 (0.0\%) & 2 (0.3\%) & 7 (4.1\%) \\ 
  $>$30 & 0 (0.0\%) & 0 (0.0\%) & 2 (1.2\%) \\ 
   \hline
    & \multicolumn{3}{c}{Cardiovascular disease present} \\
  \hline
  Homocystine group ($\mu$mol/L) & & & \\
$<$9 & 92 (13.8\%) & 87 (14.0\%) & 17 (9.9\%) \\ 
  9–14.99 & 180 (27.1\%) & 182 (29.3\%) & 39 (22.7\%) \\ 
  15–20 & 29 (4.4\%) & 23 (3.7\%) & 12 (7.0\%) \\ 
  20–30 & 12 (1.8\%) & 11 (1.8\%) & 14 (8.1\%) \\ 
  $>$30 & 0 (0.0\%) & 4 (0.6\%) & 9 (5.2\%) \\ 
   \hline
\end{tabular}
\end{table}

In this example, where $X$ is the homocystine level and $Y$ is cardiovascular disease, we are interested in the risk difference $p\{Y(X \geq 20) = 1\} - p\{Y(X < 9) = 1\}$. Based on the data in Table \ref{menddat}, we first consider the coarsened exposure with three levels defined by $<$9, 9-20, and $\geq 20$. Again, regardless of which set of assumptions we make regarding the third level being well-defining or not or contaminated or not, we get the bounds $-0.62$ to $0.81$ with 95\% confidence intervals $-0.67$ to $0.83$. Instead if we consider the coarsened exposure with four levels defined by $<$9, 9-14.99, 15-20, and $\geq 20$, we get the exact same numeric bounds $-0.62$ to $0.81$. In this case, since we only have summary statistics, we use the parametric bootstrap by resampling from multinomials with the observed distribution of probabilities within each level of the instrument.

\section{Discussion}
We have clarified the results of \citet{balke1995thesis} by giving clear definitions of well- and ill-defined. We have also shown that although an ill-defining coarsened level of the exposure is not inherently the same as an improper IV for that exposure level, a partially contaminated IV and an ill-defining level provide the same information for partial identification in the setting of a two-level IV. We have extended these results to a three-level IV for use in Mendelian randomization studies. We have also considered settings where the coarsened exposure may only have one well-defining level. 

We have assumed that a coarsened exposure $X$ has at least some well-defining levels; this has two immediate implications. First, all results for well-defining levels can equally apply to a measured exposure $X^*$ that happens to have that same number of well-defining levels. Second, when there are no well-defining levels, symbolic bounds, at least the ones derived using the method of \citet{sachs2022general}, cannot be applied. Instead, there are several methods that are conceptually different, which treat the realized exposure as a distribution; of particular note is the representative interventions literature, specifically \citet{young2018inverse}. 

There is a notable pattern in the novel and previously derived bounds; particularly, the bounds for a well-defining coarsened exposure with two or three levels with or without the addition of a potentially ill-defining level are symbolically identical. Additionally, for a two-level IV, the bounds for the risk difference with well-defining two-level exposure are contained in the bounds for a well-defining three-level exposure, and the bounds for the counterfactual risk given only one well-defining level is contained in the bounds for a well-defining two-level exposure. Although it is difficult to determine analytically if this is also the case for the increasing levels of the exposure for a three-level IV, the results from the real data example suggest this is likely the case. Investigating if this pattern extends to higher levels of the exposure and if so, if it can be proven for arbitrary levels is an area of ongoing research.

\bibliographystyle{plainnat}
\bibliography{main}

\end{document}